\documentstyle[11pt,aaspp4,tighten]{article}

\begin{document}

\title{A Comparative Study of the Parker Instability under Three Models
of the Galactic Gravity\altaffilmark{5}}
\author{Jongsoo Kim\altaffilmark{1,2,3},
    and S. S. Hong\altaffilmark{2,4}}

\altaffiltext{1}
{Korea Astronomy Observatory, San 36-1, Hwaam-Dong, Yusong-Ku,
Taejon 305-348, Korea}
\altaffiltext{2}
{Department of Astronomy, Seoul National University, Seoul 151-742, Korea}
\altaffiltext{3}
{e-mail: jskim@hanul.issa.re.kr}
\altaffiltext{4}
{e-mail: sshong@astroism.snu.ac.kr}
\altaffiltext{5}
{To appear in the Astrophysical Journal, November 1, 1998 issue, Vol. 507}

\begin{abstract}
To examine how non-uniform nature of the Galactic gravity might 
affect length and time scales of the Parker instability, we took 
three models of gravity, including the usual uniform one. In a linear 
model we let the acceleration perpendicular to the Galactic plane 
increase linearly with vertical distance $z$ from the mid-plane. 
As a more realistic choice, we let a hyperbolic tangent function of 
$z$ describe the observationally known variation of the vertical 
acceleration. To make comparisons of the three gravity models on a 
common basis, we first fixed the ratio of magnetic pressure to gas 
pressure at $\alpha$ = 0.25, that of cosmic-ray pressure at $\beta$ 
= 0.4, and the {\it rms} velocity of interstellar clouds at $a_s$ 
= 6.4 km s$^{-1}$, and then adjusted parameters of the gravity models 
in such a way that the resulting density scale heights for the three 
models may all have the same value of 160 pc.   

In the initial equilibrium state, the vertical density structure is given by an 
exponential, Gaussian, and power of hyperbolic cosine functions of $z$ 
for the uniform, linear, and realistic gravity models, respectively. 
Performing linear stability analyses onto these equilibria with the 
same ISM conditions specified by the above $\alpha$, $\beta$, and $a_s$ 
values, we calculate the maximum growth rate and corresponding length 
scale for each of the gravity models. Under the uniform gravity the Parker 
instability has the growth time of 1.2$\times$10$^{8}$ years and the 
length scale of 1.6 kpc for symmetric mode. Under the realistic gravity 
it grows in 1.8$\times$10$^{7}$ years for both symmetric and antisymmetric 
modes, and develops density condensations at intervals of 400 pc for the 
symmetric mode and 200 pc for the antisymmetric one. A simple change of 
the gravity model has thus reduced the growth time by almost an order of 
magnitude and its length scale by factors of four to eight. These results 
suggest that an onset of the Parker instability in the ISM may not 
necessarily be confined to the regions of high $\alpha$ and $\beta$.
\end{abstract}

\keywords{instabilities --- ISM: clouds --- ISM: magnetic fields
--- magnetohydrodynamics: MHD}

\section{INTRODUCTION} 

By performing linear stability analysis upon an initial equilibrium state 
of interstellar matter (ISM) that is supported against an externally given 
uniform gravity by the pressures of interstellar gas, magnetic fields, 
and cosmic ray particles, Parker(1966) proved that such a system is 
unstable to long wavelength perturbations along the direction of initial 
unperturbed magnetic fields. Since then many studies investigated effects 
on the Parker instability of rotation (Shu 1974; Zweibel \& Kulsrud 1975; 
Foglizzo \& Tagger 1994), magnetic microturbulence (Zweibel \& Kulsrud 
1975), skewed configuration of magnetic fields (Hanawa {\it et al}. 
1992a), non-uniform nature of the externally given gravity (Horiuchi 
{\it et al}. 1988; Giz \& Shu 1993; Kim {\it et al}. 1997), self-gravity 
of the ISM (Elmegreen 1982; Nakamura {\it et al}. 1991; Hanawa {\it et 
al}. 1992b), and the Galactic corona (Kamaya {\it et al}. 1997). The 
Parker instability is expected to play significant roles in the dynamo 
action of accretion disks (Tout \& Pringle 1992) and the ejection of mass 
to galactic halos (Kamaya {\it et al}. 1996). The instability has been 
thought an important formation mechanism of the giant molecular cloud 
complexes (GMCs) in the Galaxy (Appenzeller 1974; Mouschovias {\it et al}. 
1974; Blitz \& Shu 1980; Shibata \& Matsumoto 1991; Handa {\it et al}. 
1992; Gomez de Castro \& Pudritz 1992). 

As a formation mechanism of the GMCs, the Parker instability under the 
uniform gravity is facing two severe problems. The fastest growing mode 
has an infinite wavenumber along the radial direction, which is the 
direction perpendicular to both the unperturbed magnetic field and the 
externally given gravity (Parker 1967). Therefore, small-scale chaotic 
structures rather than large-scale condensations are likely to form 
through the instability (Ass\'{e}o {\it et al}. 1980; Kim {\it et al}. 
1998). Another problem lies in the time and length scales. If 
perturbations of finite vertical wavelength are given, at canonical 
ISM conditions ({\it cf}. Spitzer 1978), the Parker instability under 
the uniform gravity grows in a time scale of 1.2$\times$10$^{8}$ years 
with the corresponding length scale being 1.6 kpc. These scales are too 
long and too large for the Parker instability to be the formation 
mechanism of the GMCs, since lifetime of interstellar clouds is only 
about 3$\times$10$^{7}$ years (Blitz \& Shu 1980) and mean 
separation of the GMCs is observed to be about 0.5 kpc (Blitz 1991).

In relation to the second problem of scales, Mouschovias {\it et al}. 
(1974) pointed out that physical conditions behind the Galactic shocks 
are more favorable to trigger the Parker instability than in general 
interstellar space. It is true that in the shocked region one may reduce 
both scales by significant factors. But at the same time an increase in 
density there makes heretofore ignored self-gravity important. The Jeans 
instability would then override the Parker instability (Elmegreen 1982).

The first problem of an infinite wavenumber has its root at the
Rayleigh-Taylor 
instability, whose growth rate increases with increasing wavenumber. 
Hanawa {\it et al}. (1992a) sought a solution to this problem of infinite 
wavenumber from a skewed magnetic field. If the field lines change their 
directions systematically with height from the Galactic mid-plane, the 
buoyancy is likely to be suppressed. According to their analysis, the 
fastest growing mode of the Parker instability takes a wavelength of 
$\lambda\simeq$10$H$, with $H$ being the density scale height. As long 
as the skewness is appreciable, for example, more than 30$^{\rm o}$/$H$, 
the wavelength of maximum growth rate doesn't seem to depend sensitively 
on the degree of skewness. Observational confirmations are still needed 
on the skewness of the Galactic magnetic fields. Even if a significant 
skewness is confirmed, the resulting scale length of 10$H$ is again too 
large for the GMCs. This brings us back to the second problem of scales 
involved in the perturbations along the field line.

Parker (1966) assumed, in his pioneering study, the vertical gravitational 
acceleration to be a constant of $z$. In many ensuing studies the assumption 
of uniform gravity was almost exclusively employed in estimating the time 
and length scales of the instability. The values of 1.2$\times$10$^{8}$ 
years and 1.6 kpc we quoted above and $\lambda\simeq$10$H$ of Hanawa {\it 
et al}. (1992a) are all based on the same assumption of uniform gravity. 
However, we do know that the Galactic gravity is of non-uniform nature 
(Oort 1965; Bienaym\'{e}, Robin \& Cr\'{e}z\'{e} 1987, hereafter BRC). 
It is then interesting to see how much reductions one may achieve in the 
length and time scales by introducing a realistic model for the Galactic 
gravity.  

The uniform gravity assumption has brought a few unphysical features 
to the classical picture of the Parker instability. Under a constant 
acceleration, the equilibrium distribution of density follows an
exponential function of height $z$  and has a cusp at the mid-plane, 
which accompanies a discontinuity in the pressure gradient at $z$ = 0. 
Because of the discontinuity the linear stability analyses based on the 
uniform gravity ought to be limited to the perturbations of mirror 
symmetric mode only; while the perturbations of antisymmetric mode are 
expected to dominate the symmetric ones (Horiuchi {\it et al}. 1988; 
Basu {\it et al}. 1997). The antisymmetric mode would not only slightly 
increase the growth rate over the value of mirror symmetry, but also 
obviously reduce the distance between condensations by a factor of two. 
Therefore, models other than the uniform gravity are required to make 
better estimations of the time and length scales for the Parker 
instability in the Galaxy. 

In their linear stability analysis, Giz \& Shu(1993) introduced a 
hyperbolic tangent function of {\it z} to model the vertical variation 
of the Galactic gravity. And recently Kim {\it et al}. (1997) used a 
linear function of {\it z} as a non-uniform model of the Galactic 
gravity. Both studies demonstrated that dynamical evolution of the 
perturbations given to the Galactic disk would follow, under such 
non-uniform gravities, two different families of solutions. They 
are called {\it continuum} and {\it discrete} families. The familiar 
continuum family has two modes of solutions: One is the original Parker 
mode, which is a slow magnetohydrodynamic (MHD) mode modulated by the 
gravity, and the other is a stable Alfv\'{e}n mode. The newly discovered 
discrete family has three modes: stable fast MHD, stable slow MHD, and 
{\it unstable} slow MHD modes. Both studies put 
their emphases on the discovery and characterization of the discrete 
family, and didn't fully address the question of GMC formation in the 
context of non-uniform gravity. When an effective adiabatic index of 
the ISM is larger than a certain critical value, only the discrete 
family can have unstable solutions (Giz \& Shu 1993; Kim {\it et al}. 
1997). However, the growth time of the unstable {\it discrete} solution 
turned out to be about 10 times longer than that of the unstable {\it 
continuum} solution. It is, therefore, difficult to explain formation 
of the GMCs by the discrete family of solutions. 

In the present study we will concentrate on the continuum family of 
solutions. Since the external gravity is the driving force of the Parker
instability, its time and length scales should depend on the nature of 
the Galactic gravity. This line of reasonings motivated us to seek a 
solution to the second problem of scales from the non-uniform nature of 
the Galactic gravity. 

This paper is organized as follows. In \S 2 we first introduce three 
models for the Galactic gravity, and then construct equilibrium 
configuration for each of the gravity models. The MHD equations are 
linearized in the same section. In \S 3 dispersion relations are derived 
for the three gravity models, and the resulting sets of time and length 
scales are compared with each other. In \S 4 we summarize the paper with 
some discussions.

\section{FORMULATION}

Complete MHD equations for the magnetized gas and cosmic-ray particles
under an externally given gravitational field are given by
\begin{equation}
     {\partial \rho \over \partial t} +
     \nabla \cdot (\rho \vec v) =0,
\end{equation}
\begin{equation}
\rho
\left[ {\partial \vec v \over \partial t}
+ (\vec v \cdot \nabla)\vec v \right]
= -\nabla \left(p + P +{B^2 \over 8\pi}\right)
+{1 \over 4\pi}\vec B \cdot \nabla \vec B
+ \rho \vec g,
\end{equation}
\begin{equation}
 {\partial \vec B \over \partial t} =
      \nabla \times (\vec v \times \vec B),
\end{equation}
\begin{equation}
{\partial p \over \partial t} + \vec v \cdot \nabla p
 + \gamma p \nabla \cdot \vec v =0,
\end{equation}
\begin{equation}
 \vec B \cdot \nabla P = 0.
\end{equation}
Here gas and cosmic-ray pressures are denoted by $p$ and $P$, 
respectively. Other symbols have their usual meanings. Since the 
dispersion of random thermal velocities is much smaller than that of 
random cloud velocities, the main source of gas pressure is from the 
macroscopic turbulent motion of the clouds. To describe the local 
behavior of the Parker instability in the Galactic disk we introduce 
Cartesian coordinates $x$, $y$, and $z$, whose axes are taken to be 
parallel to the radial, azimuthal and vertical directions, respectively. 
The gravity has only vertical component, whose magnitude varies with 
height $z$, {\it i.e.} $\vec g = [0,0,-g(z)]$. 

If $\gamma$ is larger than one and smaller than a critical value, 
the system becomes unstable by both families of continuum and discrete; 
if $\gamma$ is less than one, there exists convective instability 
(Kim {\it et al}. 1997). Since our principal concern is to see the 
effects of gravity on the Parker instability, we simply choose 
$\gamma$ = 1 and limit ourselves to the unstable continuum family of 
solutions. 
 
\subsection{Three Models for the Galactic Gravity}

In the Galaxy the main source of gravity is stars, and the vertical 
gravity can be represented by
\begin{equation}\label{rg}
g_r(z) = 2 \frac{<\!\! v_*^2 \!\!>}{H_*} \tanh\left(\frac{z}{H_*}\right),
\end{equation}
where $<\!\! v_*^2 \!\!>$ and $H_*$ are the velocity dispersion and the  
scale height of the stars, respectively (Kim 1990; Giz \& Shu 1993).  
We call this a realistic gravity as Giz \& Shu did.  In addition to 
this, we consider a uniform gravity,
\begin{equation}\label{ug}
g_u(z) = g_o \frac{z}{|z|},
\end{equation}
and a linear gravity,
\begin{equation}\label{lg}
g_l(z) = g'z,
\end{equation}
where both $g_o$ and $g'$ are positive constants.

The uniform, linear and realistic gravity models are represented in 
Figure~1 by the dotted, dashed, and long-dashed lines, respectively.  
The two solid lines in the figure represent the strength of the 
gravitational accelerations of Oort (1965) and of BRC at solar 
neighborhood as functions of $z$. The gravity inferred from the 
distribution of K giants (Oort 1965) increases almost linearly up 
to $z \sim 500$ pc, beyond which it stays more-or-less constant.  
The same trend can be found from the gravity based on a galaxy 
model (BRC).

\subsection{Unperturbed States}

Let us suppose that initially an infinitely extended disk of gas and 
cosmic-ray particles is under the influences of magnetic and gravitational 
fields. The unperturbed magnetic field $\vec B_o$ has only an azimuthal 
component, whose magnitude varies with $z$, {\it i.e.} $\vec B_o = 
[0,B_o(z),0]$. Then the magnetohydrostatic equilibrium of the system is 
governed by
\begin{equation}
   {d \over dz} \left[ p_o(z) + P_o(z) + {B_o^2(z) \over 8 \pi} \right] =
     - \rho_o(z) g(z).
\end{equation}
To close the differential equation, we take an 
isothermal equation of state for the  gas pressure $p_o=a_s^2\rho_o(z)$, 
where the velocity dispersion of clouds $a_s^2$ is assumed constant 
everywhere. We also assume that the ratio $\alpha = B_o^2/8\pi p_o$ of 
magnetic pressure to gas pressure and the ratio $\beta = P_o/p_o$ of 
cosmic-ray pressure to gas pressure are constants.  Then, we have the 
same $z$-dependence for the unperturbed distributions of density, gas 
pressure, cosmic-ray pressure, and magnetic pressure as
\begin{equation}\label{us} 
{\rho_o(z) \over \rho_o(0)} =
 {p_o(z) \over p_o(0)} =
 {P_o(z) \over P_o(0)} =
 {B_o^2(z) \over B_o^2(0)} =
 \exp  \left[ - {\Phi(z) \over (1 + \alpha + \beta)a_s^2} \right],
\end{equation}
where  $\Phi(z)$ is defined by
\begin{equation}\label{gp}
        \Phi(z) \equiv  \int^z_0 g(z) dz,
\end{equation}
and $\rho_o(0)$, $p_o(0)$, $P_o(0)$, and $B_o(0)$ 
denote their mid-plane values. An effective scale height $H$ of the 
density distribution is defined by
\begin{equation}\label{H}
2H \equiv {1 \over \rho_o(0)} \int^{+\infty}_{-\infty} \rho_o(z) dz.
\end{equation}

Under the uniform gravity, the unperturbed state is described by 
an exponential function,
\begin{equation}\label{usu}
 {\rho_{o,u}(z) \over \rho_{o,u}(0)} =
 {p_{o,u}(z) \over p_{o,u}(0)} =
 {P_{o,u}(z) \over P_{o,u}(0)} =
 {B_{o,u}^2(z) \over B_{o,u}^2(0)} =
 \exp  \left[ - {|z| \over H} \right],
\end{equation}
with the scale height being
\begin{equation}\label{Hu}
  H (\equiv H_u) = {(1+ \alpha + \beta)a_s^2 \over g_o}.
\end{equation}
Under the linear gravity, the unperturbed state is given by a Gaussian 
function,
\begin{equation}\label{usl}
 {\rho_{o,l}(z) \over \rho_{o,l}(0)} =
 {p_{o,l}(z) \over p_{o,l}(0)} =
 {P_{o,l}(z) \over P_{o,l}(0)} =
 {B_{o,l}^2(z) \over B_{o,l}^2(0)} =
 \exp  \left[ - {\pi \over      
  4}\left({z \over H}\right)^2 \right],
\end{equation}
where the scale height now becomes
\begin{equation}\label{Hl}
    H (\equiv H_l) =  
    a_s \left[ {\pi(1+\alpha+\beta) \over {2g'}} \right]^{{1 \over 2}}.
\end{equation}
For the case of the realistic gravity, it is described by
\begin{equation}\label{upr}
 {\rho_{o,r}(z) \over \rho_{o,r}(0)} = 
 {p_{o,r}(z) \over p_{o,r}(0)} =
 {P_{o,r}(z) \over P_{o,r}(0)} =
 {B_{o,r}^2(z) \over B_{o,r}^2(0)} =
 {\rm sech}^{2s} \left( {z \over H_*} \right),
\end{equation}
where $s$ is defined by
\begin{equation}\label{s}
s \equiv \frac{<\!\! v_*^2 \!\!>}{(1+\alpha+\beta)a_s^2}.
\end{equation}
If we call $(1+\alpha+\beta)a_s^2$ as an {\it effective velocity 
dispersion} of clouds, then $s$ is the ratio of the velocity dispersion 
of stars to the effective velocity dispersion of clouds. Substituting 
equation~(\ref{upr}) into equation~({\ref{H}), we may describe 
the scale height of clouds in terms of $H_*$ and $s$,
\begin{equation}\label{Hr}
   H (\equiv H_r) = \frac{(2s-2)!!}{(2s-1)!!} H_* \equiv \frac{H_*}{h},
\end{equation}
where $h$ is the ratio of the scale height of stars to that of clouds.

In order to specify the gravity models one should fix all the parameters
$g_o$, $g'$, $H_*$, and $<\!\! v_*^2 \!\!>$.  Under each model of 
gravity, the magnetized gas and cosmic-ray particles adjust themselves 
to the equilibrium state.  This means that the gravity parameters
are related to the ISM parameters, $\alpha$, $\beta$, $H$, and $a_s$
(see eq.~[14] for uniform gravity, eq.~[16] for linear gravity, and
eqs. [18] and [19] for realistic gravity). In practice, however,
it is difficult to uniquely determine the values of $g_o$ and $g'$ from
either Oort's (1965) or BRC's gravity.  Furthermore, the values
of $H_*$ and $<\!\! v_*^2 \!\!>$ for one type of stars are different
from those for other type of stars (Mihalas \& Binney 1981). Therefore 
we decide to pin down the parameter values on the basis of the ISM 
conditions. They are specified in such a way that the gas scale height
resulted from each of the three gravity models may all take the same
value 160 pc that is known from observations (Falgarone \& Lequeux 1973).
In this way the comparison of the length and time scales
among the three gravity models can be done on a common ground. 

We take $\alpha=0.25$ and $\beta=0.4$, which are canonical values of the 
ISM (Spitzer 1978). For the scale height and the {\it rms} velocity 
of interstellar clouds we take 160 pc and 6.4 km s$^{-1}$, respectively 
(Falgarone \& Lequeux 1973). Then, $g_o$ and $g'$ are equal to 
1.4$\times$10$^{-9}$ cm s$^{-2}$, and 13.4$\times$10$^{-9}$ cm s$^{-2}$ 
kpc$^{-1}$. In addition to the above values of $\alpha$, $\beta$, $H$, 
and $a_s$, we should also fix $H_*$ and $<\!\! v_*^2 \!\!>$ according to 
the chosen value of parameter $s$. This completes the specification of 
the realistic gravity model. The resulting three models are compared in 
Figure~1.

The parameter $s$ defined by equation~(18) is exactly the same as $R$ in Giz 
\& Shu (1993). They took $R \approx 3.5$ by fixing the value of 
$R_{\rm eq} \approx 2$, ratio of the equivalent half-thickness of 
stellar disk to gas disk.  But the equivalent half-thicknesses for 
various stellar objects in the Galaxy are different from type to type. 
In this paper, however, we will fix the $s$ value in such a way that the 
resulting model of gravity may closely resemble the gravities of Oort 
(1965) and BRC.  As can be seen from Figure~1, their gravities are 
reproduced by the realistic model with $s=16$ and $s=9$, respectively. 

\subsection{Linearized Perturbation Equations}

We limit ourselves to the perturbations that are propagating in the plane 
defined by the azimuthal and vertical directions.  In the two dimensional 
geometry the magnetic vector potential $\vec A=A(y,z)\hat x$ is more 
convenient to use than the magnetic field $\vec B$, because one scalar 
quantity, $A(y,z)$, is enough to specify the $y$ and $z$ components of 
the field. Taking the advantage we combine the linearized equations ~(1) 
through (5) into one for $\delta A$: 
\begin{eqnarray}
&&-Q^2{\partial^2\over\partial t^2}\delta A
 + a_s^2\left[(2\alpha+\gamma)Q^2
 + \gamma^2a_s^2{\partial^2\over\partial y^2}\right]
 {\partial^2\over\partial z^2}\delta A \nonumber \\ 
&&+\left\{ a_s^2 \left[ 2\alpha {\partial^2\over\partial y^2}
 +\left(1+\beta - {\gamma \over 2}\right){d^2\over dz^2}\ln\rho_o
 -{1 \over 2}\left(\alpha + {\gamma \over 2}\right)
 \left({d\over dz}\ln\rho_o\right)^2 \right] Q^2 \right. 
\label{pe} \\
&&\left.+a_s^4 \left[\gamma\left(1+\alpha+\beta-{\gamma\over 2}\right)
 {d^2 \over dz^2}\ln\rho_o-\left(1+\alpha+\beta-{\gamma\over 2}\right)^2
 \left({d\over dz}\ln\rho_o\right)^2 \right]{\partial ^2\over\partial y^2}
 \right\}\delta A = 0, \nonumber 
\end{eqnarray}
where $Q^2$ is an acoustic wave operator defined by
\begin{equation}
  Q^2 \equiv {\partial ^2 \over \partial t^2} - \gamma a_s^2
        {\partial ^2 \over \partial y^2}.
\end{equation}
If the following function
\begin{equation}\label{dA}
\delta A = f(z) \exp (i\omega t - ik_yy)
\end{equation}
is substituted for the perturbation of vector potential, equation 
(\ref{pe}) becomes a second order ordinary differential equation  
\begin{eqnarray}
a_s^2 \left[ (2\alpha+\gamma)\omega^2-2\alpha\gamma a_s^2 k_y^2 \right]
\frac{d^2f}{dz^2} \nonumber \\
+\left\{ \omega^4 - a_s^2\left[(2\alpha+\gamma)k_y^2 
                  + \frac{1}{2} \left( \alpha+\frac{\gamma}{2} \right)
                    \left( \frac{d}{dz}\ln\rho_o\right)^2
                  - \left( 1+\beta-\frac{\gamma}{2} \right)
                    \left( \frac{d^2}{dz^2} \ln \rho_o \right) \right] 
\omega^2 \right\} f(z) \\
+a_s^4 k_y^2 \left\{ 2\alpha\gamma k_y^2
                  -\left[ (1+\alpha+\beta)(1+\alpha+\beta-\gamma)
                          -\frac{1}{2}\alpha\gamma \right]
                   \left( \frac{d}{dz} \ln \rho_o \right)^2 
                  +\alpha\gamma 
                   \left(\frac{d^2}{dz^2} \ln \rho_o \right) \right\}
                  f(z) = 0, \nonumber
\end{eqnarray}
where $f(z)$ is an amplitude function of the perturbation, $\omega$ is 
the angular frequency, and $k_y$ is the wavenumber along the azimuthal 
direction. 

As an upper boundary condition (BC), one may set $f=0$ at 
$z=z_{\rm node}$. The value of $z_{\rm node}$ can be either finite or 
infinite. As a lower BC, one may take either $f=0$ or $df/dz = 0$ at 
$z=0$. The former generates solutions having mirror symmetry; while the 
latter does the ones having antisymmetry. Since $z$-distribution of the 
unperturbed state has a cusp at the mid-plane under the uniform gravity 
(eq. [\ref{usu}]), only one condition, $f=0$, is applicable to the lower 
boundary for the case of uniform gravity.  

The effective scale height $H$ and the crossing time over one scale 
height $H/a_s$ are taken as normalization units of length and 
time. Dimensionless variables are then defined by
\begin{equation}
\Omega \equiv i\omega H/a_s,         \;\;
\zeta \equiv z/H,                     \;\;
\nu_y \equiv k_yH.                   \;\;
\end{equation}
The dimensionless vertical wavenumber, which depends on the lower BCs, 
should be defined by
\begin{equation}
\nu_z \equiv \left\{ \begin{array}{ll}
                       2\pi(H/2z_{\rm node}) & \mbox{for symmetric 
modes} \\  
                       2\pi(H/4z_{\rm node}) & \mbox{for antisymmetric 
modes}.  
                     \end{array}
             \right.
\end{equation}
Since we used the same scale height $H$ and the same {\it rms} velocity 
$a_s$ for all the models, it is not necessary to distinguish the 
dimensionless variables from model to model.

\section{TIME AND LENGTH SCALES}
 
\subsection{Uniform Gravity}

In this paper we will re-derive dispersion relations for the Parker 
instability under the uniform gravity, because we want to point out the 
problems involved in the time and the length scales, and because the 
results from the uniform model comprise a comparison basis. Under the 
uniform gravity $d \ln \rho_o(z)/dz$ and $d^2 \ln \rho_o(z) / dz^2$ in 
equation (23) should be replaced by $-1/H$ and $0$, respectively 
(see eq.~[\ref{usu}]). In terms of the dimensionless variables, it 
takes the form  
\begin{eqnarray}
&&\left[(2\alpha+\gamma)\Omega^2+2\alpha\gamma\nu_y^2\right]
   {d^2f_u\over d\zeta^2} 
   +\left\{-\Omega^4 
   -(2\alpha+\gamma)\left(\nu_y^2+\frac{1}{4}\right)\Omega^2
   \right. \nonumber \\
&& \left.   
          -2\alpha\gamma\nu_y^4 
 +\left[(1+\alpha + \beta)(1+\alpha+\beta-\gamma)
        -{\frac{1}{2}\alpha\gamma}\right]\nu_y^2\right\}f_u=0,  
\end{eqnarray}
which is what Parker(1966) gave in his Appendix III.  Because the 
coefficients of $d^2f_u/d\zeta^2$ and $f_u$ are constants, and because 
only symmetric BC at $\zeta=0$ should be applied to the model of the 
uniform gravity, we may set $f_u \propto \sin(\nu_z\zeta)$. Then, an 
equation for the dispersion relation can be written
\begin{eqnarray}
&& \Omega^4
  +(2\alpha+\gamma)\left(\nu_y^2+\nu_z^2+\frac{1}{4}\right) \Omega^2 
\nonumber \\
&&+\left\{2\alpha\gamma(\nu_y^2+\nu_z^2) 
    -\left[(1+\alpha+\beta)(1+\alpha+\beta-\gamma)-\frac{1}{2}\alpha\gamma
     \right]
   \right\} \nu_y^2 = 0.
\end{eqnarray}

The resulting dispersion relations are shown in Figure 2. The perturbation 
with a finite vertical wavelength grows less rapidly than the one with an 
infinite wavelength. If we take $H$ = 160 pc, $\nu_z$ = 1.0 corresponds to 
$z_{\rm node}\simeq$ 500 pc. Since the scale heights of interstellar 
clouds and inter-cloud gas are 160 pc and 300 pc (Falgarone and Lequeux 
1973), respectively, 500 pc is a reasonable choice for the nodal point 
(e.g., Elmegreen 1982). For the perturbation with the infinite vertical 
wavelength, the minimum growth time of our result is 5.5$\times$10$^7$ 
years and Parker's (1966) estimate is 3$\times$10$^7$ years. Besides 
detailed values of $\alpha$ and $\beta$ that went into these estimates, 
the time scale of 3$\sim$6$\times$10$^7$ years is a gross under-estimate; 
under the Galactic environments it seems unrealistic to think of 
perturbations whose vertical wavelength is much larger than the scale 
height of the cloud distribution itself. In the case of $\nu_z$ = 1.0, 
the time and length scales become 1.2$\times$10$^8$ years and 1.6 kpc, 
respectively. The growth time, 1.2$\times$10$^8$ years, is much longer 
than the lifetime of GMCs, 3$\times$10$^7$ years (Blitz and Shu 1980), 
and the length scale, 1.6 kpc, is larger than the mean separation of 
the GMCs, 0.5 kpc (Blitz 1991). 

Figure~3 illustrates the detailed dependences of the the minimum growth 
time upon the parameters $\alpha$ and $\beta$. In most region of the 
($\alpha, \beta$) plane the growth time turns out to be longer than the 
cloud lifetime. From the illustration we conclude that under the uniform 
gravity the Parker instability may not play any significant roles in the 
formation of GMCs. The growth time becomes comparable to the lifetime of 
GMCs only in a limited region of the parameter plane, where $\alpha$ and 
$\beta$ take unacceptably high values for the general ISM. This is exactly 
the reason why Mouschovias {\it et al}. (1974) invoked the dense region of 
Galactic shocks as an onset place of the Parker instability. 

\subsection{Linear Gravity}

Under the linear gravity, $d \ln \rho_{o,l}(z) /dz$ becomes $-(\pi/2)
(z/H^2)$ (see eq. [\ref{usl}]).  After substituting it and its gradient 
into equation (23), we can write the resulting equation in the following 
dimensionless form:
\begin{equation}
\frac{d^2 f_l}{d\zeta^2} + (E_l -V_{o,l} \ \zeta^2) f_l =0,
\end{equation}
where $E_l$ and $V_{o,l}$ are given by
\begin{equation}
E_l = \frac{\pi}{4} 
    - \frac{\Omega^4 + [(2\alpha+\gamma)\nu_y^2
                         +\frac{\pi}{2}(1+\alpha+\beta)]\Omega^2
                     +2\alpha\gamma\nu_y^4}
           {(2\alpha+\gamma)\Omega^2+2\alpha\gamma\nu_y^2},
\end{equation}
\begin{equation}
V_{o,l} = \frac{\pi^2}{4} 
        \left[ \frac{1}{4} - 
               \frac{(1+\alpha+\beta)(1+\alpha+\beta-\gamma)\nu_y^2}
                    {(2\alpha+\gamma)\Omega^2+2\alpha\gamma\nu_y^2}
        \right].
\end{equation}
To derive the dispersion relation we will use the same method given 
in the Appendix of Kim {\it et al}. (1997).

Kim {\it et al}. (1997) showed that, under the linear gravity, the 
dispersion relation for the continuum family of solutions with 
a mirror symmetric lower BC are nearly the same as that with an 
antisymmetric BC. Under the point-mass-dominated gravity, however, 
the antisymmetric modes grow faster than the symmetric ones (Horiuchi 
{\it et al}. 1988). Such difference in the behavior of growth rate 
stems from the difference in the nature of the chosen gravity models. 
The growth rate under the point-mass-dominated gravity is sensitive 
to the lower BCs, because the gravity has its maximum close to the 
lower boundary. On the contrary the growth rates under the linear 
and realistic gravities are insensitive to the lower BCs, because 
these models have their maxima at far from the mid-plane. So in this 
paper we present the dispersion relations only for the symmetric 
lower BC.

In the linear model the vertical acceleration increases with $z$ without 
a bound, while in the Galaxy it approaches a finite value at large distance 
from the mid-plane. We should, therefore, limit the disk extent within 
a finite height from the central plane. The dispersion relations shown 
in Figure~4 are for the perturbations with $\nu_z = 0.5$ and $\nu_z = 
1.0$. This figure clearly indicates a strong dependence of the growth rate 
on the vertical wavenumber. We think $\nu_z = 1.0$ a realistic choice for 
the Galaxy. With the same parameters that are used in Figure~2, the 
perturbation of maximum growth rate has a horizontal wavelength 340 pc 
and a minimum growth time 1.5$\times$10$^7$ years.

The linear gravity drives the Parker instability much faster than the 
uniform gravity. By almost an order of magnitude reduction is achieved 
in the growth time scale, even with the perturbation wavelength as short 
as 340 pc.  Such reduction is possible for a wide range of $\alpha$ and 
$\beta$ values. This can be seen from the contours, in Figure~5, of equal 
minimum growth times traced out in the $(\alpha, \beta)$ plane. Comparison 
of the two sets of figures (Figs.~2 and 3 versus Figs.~4 and 5) indicates 
that the growth rate is indeed sensitive to the nature of the externally 
given gravitational fields. 

\subsection{Realistic Gravity}

Using the unperturbed state expressed by equation~(\ref{upr}) together
with the definition of $h$ (eq. [\ref{Hr}]), we may express the 
perturbation equation~(23) in the form 
\begin{equation}
\frac{d^2f_r}{d\zeta^2} +
\left[ E_r - V_{o,r} \ {\rm sech}^2 \left( \frac{\zeta}{h} \right) 
\right] f_r = 0,
\end{equation}
with
\begin{equation}
E_r = (G_3+G_4) / G_1, 
\end{equation}
\begin{equation}
V_{o,r} = (G_3-G_2) / G_1,
\end{equation}
where $G_1$, $G_2$, $G_3$, and $G_4$ are given by
\begin{equation}
G_1 = (2\alpha+\gamma)\Omega^2 + 2\alpha\gamma\nu_y^2,
\end{equation}
\begin{equation}
G_2 = \left[ -(2+2\beta-\gamma)\Omega^2 + 2\alpha\gamma\nu_y^2\right]
          \frac{s}{h^2},
\end{equation}
\begin{equation}
G_3 = \left\{ -(2\alpha+\gamma)\Omega^2 
            +[4(1+\alpha+\beta)(1+\alpha+\beta-\gamma)-2\alpha\gamma]
\nu_y^2\right\} \frac{s^2}{h^2},
\end{equation}
\begin{equation}
G_4 = -\Omega^4 -(2\alpha+\gamma)\nu_y^2\Omega^2 - 2\alpha\gamma
\nu_y^4.
\end{equation}
To evaluate the dispersion relations we again use the method explained 
in the Appendix of Kim {\it et al}. (1997). 

The resulting dispersion relations of the symmetric mode are shown in 
Figure~6 for different values of $s$. The larger the $s$ value is, the 
more unstable the system becomes. This is because the gravity with 
a larger value of $s$ is stronger than that with a smaller one (see 
Fig.~1). To show the effect of the vertical wavenumber on the growth 
rate, the dispersion relations for the cases of $\nu_z$ = 0.0, 0.5, and 
1.0 are compared to each other in Figure~7. The smaller the vertical 
wavenumber or the longer the vertical wavelength is, the faster the 
instability grows. For the  set of parameters $s=9, \nu_z=1, \alpha=0.25, 
\beta=0.4$, and $\gamma=1$, the minimum growth time becomes 
1.8$\times$10$^7$ years and the corresponding wavelength is 400 pc. 
If $s=16$ is taken with the other parameters being fixed at the same 
values, the time and length scales are slightly reduced. Figure~8 shows 
the equi-growth time contours in the $(\alpha, \beta)$ plane for the 
case of $\gamma=1, \nu_z=1$, and $s=9$. Under the realistic gravity the 
growth time becomes shorter than the GMC lifetime for most range of 
the $\alpha$ and $\beta$ values.

\subsection{Comparison of Time and Length Scales}

The dispersion relations resulting from the three models of gravity 
are compared in Figure~9. The same set of parameter values, $\alpha = 
0.25$, $\beta = 0.4$, $\gamma = 1.0$, $\nu_z = 1.0$, is used for all 
the three models. To do a good justice in the comparison, the time and 
length scales of the three models are normalized by using the same 
set of effective scale height 160 pc and {\it rms} velocity 6.4 km 
s$^{-1}$. The vertical wavenumber $\nu_z = 1$ places a nodal point 
at $z \simeq$ 500 pc.  Among the three the linear gravity gives the 
strongest acceleration near the nodal point; while the uniform gravity 
does the weakest (see Fig. 1). Since it is the external gravity that 
drives the Parker instability, the growth rates for the linear gravity 
are generally higher than for the other two.

Table~1 lists the minimum growth times and their length scales for each 
of the three gravity models. We include the scales for both the symmetric 
and antisymmetric modes. The parameters used in the calculation of these 
scales are the same as those in Figure~9. The length scale corresponds  
to the inter-distance of the condensations that would be formed by the 
Parker instability. So the length scale for the symmetric mode, which 
is equal to the perturbation wavelength, is twice the value for the 
antisymmetric one. We should also clarify the vertical wavenumber used 
for the antisymmetric modes.  As is done for the symmetric modes, the 
upper nodal point is set at $z \simeq 500$ pc for the antisymmetric ones.
Since the antisymmetric mode doesn't have a nodal point at $z=0$, this 
choice corresponds to the vertical wavenumber $\nu_z = 0.5$, which is 
smaller by a factor of two than that for the symmetric one.  It is, 
however, fair to use the scales with the same upper node rather than 
those with the same vertical wavenumber. This point should be kept in 
mind, when one compares the scales of the time and length for the 
symmetric modes with those for the antisymmetric ones. 

In spite of the incompatibility of the antisymmetric mode with the 
uniform gravity, we have listed, in Table~1, the time and length scales 
for the antisymmetric mode, because the incompatibility stems from an 
over-simplification of the Galactic gravity, and because we want to 
compare the scales from different models of gravity. Since the realistic 
gravity model with $s=9$ represents the results of BRC fairly well, we 
take 1.8$\times$10$^7$ years for the growth time scale and 400 pc (200 
pc) for its length scale of the symmetric (antisymmetric) mode of the 
Parker instability. These time and length scales are about 1/7 and 1/4 
of those for the uniform gravity, respectively.

\section{SUMMARY AND DISCUSSION}

To investigate the effect of external gravity on the continuum family 
solutions of the Parker instability, we introduce uniform, linear, 
and realistic models for the Galactic gravity. The gravity models are 
specified by fixing the ratio $\alpha$ of magnetic pressure to gas 
pressure at 0.25, the ratio $\beta$ of cosmic-ray pressure to gas 
pressure at 0.4, the {\it rms} velocity $<a_s^2>^{1/2}$ of 
interstellar clouds at 6.4 km s$^{-1}$, and the effective scale 
height $H$ of cloud distribution at 160 pc. For the realistic model 
based on the hyperbolic tangent function, we need to specify one more 
parameter $s$, which is the ratio of the velocity dispersion of stars 
to the effective velocity dispersion of clouds (see eq. [\ref{s}]). 
The Galactic gravities of Oort (1965) and BRC at solar neighborhood 
are reproduced by the realistic model with $s$ = 16 and $s$ = 9, 
respectively.

Under the uniform gravity the time and length scales of symmetric mode 
solution are 1.2$\times$10$^8$ years and 1.6 kpc; while they become 
1.8$\times$10$^7$ years and 400 pc under the realistic gravity with 
$s=9$. Under the linear and realistic gravities the antisymmetric modes 
grow at almost the same rate as the symmetric mode.  However, as for the 
separation of condensations, the length scale of the antisymmetric mode 
becomes, of course, a half of the symmetric mode's. Changing the nature 
of gravity from uniform to realistic has thus reduced the time and length 
scales in the Galactic environments by factors of 7 and 4, respectively. 
Therefore, it is not necessary to invoke high values of $\alpha$ and 
$\beta$ to reduce the scales. 

The rotation of the Galactic disk is known to exercise a stabilizing 
effect on the Parker instability (Foglizzo \& Tagger 1994). With the 
uniform gravity Zweibel \& Kulsrud (1975) found that the rigid body 
rotation would increase, over the non-rotating case, the growth time by 
a factor of two. With a gravity model based on a combination of linear 
and hyperbolic tangent functions, which is very close to ours, Hanawa 
{\it et al}. (1992b) found that the rigid body rotation would increase 
the growth time by factors of 1.3 to 7.5 for $\alpha$ varying from 1.0 
to 0.1. They placed the upper boundary of the Galactic disk at $z$ = 200 
pc, while we did at 500 pc. Because the external gravity drives the gas 
to slide down the field lines from top part of the disk first, the growth 
rate one obtains from linear stability analysis depends rather sensitively 
on the magnitude of acceleration at the upper boundary. The acceleration 
at $z$ = 200 pc is about two thirds of its value at $z$ = 500 pc. 
Therefore, an increase in the growth time the rotation would bring to 
our case may not be so large as their findings. Furthermore, they didn't 
include cosmic-ray particles as an ISM constituent. Since the  ``light'' 
cosmic-rays drive the field lines to buckle mainly upward, a triggering 
of the instability by the cosmic-ray pressure doesn't seem to be seriously 
affected by the Galactic rotation. If the Galactic rotation had been 
included in our analysis with the non-uniform gravities, the growth 
time would have increased by a factor of 2 or 3 for moderate values 
of $\alpha$ and $\beta$. However, for very low values of $\alpha$ and 
$\beta$, the rotation is very likely to make the growth time longer 
than the GMC lifetime (Figure~8).     
     
Are the GMCs formed by the Parker instability?  As far as the time and 
length scales are concerned, our linear stability analysis done with 
the realistic gravity model gives a positive answer to the question. 
To have a better answer we should also know how much enhancement in 
density can be made by the Parker instability. From the linear analysis 
alone one may not have information on the density enhancement factor. 
Very recently Basu {\it et al}. (1996; 1997) performed two-dimensional 
simulations of the Parker instability under the uniform gravity. The 
density enhancement they obtained at the mid-plane amounts to a factor 
of only 2. Under the non-uniform gravity the enhancement factor may 
not be much larger than 2, because the acceleration of any realistic 
gravity models anyhow approaches zero near the mid-plane. As for the 
density enhancement the answer is likely to be a negative one. The 
GMCs may have formed through a cooperative interplay of the Parker 
and Jeans instabilities.

\acknowledgments

We thank Dr. D. Ryu for making useful suggestions and to a referee, 
Dr. T. Foglizzo, for giving us many constructive comments. SSH wishes 
to acknowledge the financial support from the Korea Research Foundation 
made in the year of 1997. JK was supported by the Korean Ministry of 
Science and Technology through the Korea Astronomy Observatory grant 
97-5400-000.  
\clearpage

\begin{deluxetable}{ccccccc}
\tablecolumns{7}
\footnotesize
\tablenum{1}
\tablecaption{Comparison of Time and Length Scales}
\tablehead{ 
\colhead{scales} & \multicolumn{2}{c}{uniform} & 
                   \multicolumn{2}{c}{linear}  & 
                   \multicolumn{2}{c}{realistic ($s=9$)} \\
                   \cline{2-3} \cline{4-5} \cline{6-7}   \\ 
\colhead{} & \colhead{symmetric} & \colhead{antisymmetric$^*$} & 
             \colhead{symmetric} & \colhead{antisymmetric}     & 
             \colhead{symmetric} & \colhead{antisymmetric} }
\startdata
time [year]  & $1.2 \times 10^8$ & $1.2 \times 10^8$ & $1.5 \times 10^7$ 
             & $1.5 \times 10^7$ & $1.8 \times 10^7$ & $1.8 \times 10^7$ 
\nl
length [pc] & 1600 & 800 & 340 & 170 & 400 & 200 \nl

\enddata
\tablenotetext{}{$^*$Incompatible with the exponential distribution of 
the unperturbed state}

\end{deluxetable}

\clearpage

\figcaption{\label{fig1}
Vertical acceleration at solar neighborhood. The two solid lines represent 
the results of Oort (1965) and of Bienaym\'{e}, Robin \& Cr\'{e}z\'{e} 
(1987). A dotted line, a dashed line, and three long-dashed lines are for 
the models of uniform, linear, and realistic gravities, respectively. The 
parameter $s$ is the ratio of the velocity dispersion of stars to the 
effective velocity dispersion of clouds.
}  

\figcaption{\label{fig2}
Dispersion relations of the Parker instability under the uniform gravity. 
Each curve is marked by the value of vertical wavenumber. $\nu_z$ = 0.0 
corresponds to an infinite wavelength. If the density scale height of 
clouds is 160 pc, $\nu_z=0.5$ and $\nu_z$=1.0 have nodal points at 
$\simeq$ 1000 pc and $\simeq$ 500 pc, respectively. The ordinate 
represents square of the normalized growth rate, and the abscissa does 
the normalized horizontal wavenumber. The effective adiabatic index 
$\gamma$, the ratio $\alpha$ of magnetic to gas pressure, and the ratio 
$\beta$ of cosmic-ray to gas pressure are specified within the frame.
}

\figcaption{\label{fig3}
Loci of equi-growth time are traced out in the ($\alpha, \beta$) plane 
for the  Parker instability under the uniform gravity. 
}

\figcaption{\label{fig4}
Dispersion relations of the Parker instability under the linear gravity.
Each curve is marked by the vertical wavenumber $\nu_z$. The ordinate 
represents square of the normalized growth rate, and the abscissa the 
normalized horizontal wavenumber. The system parameters used in the 
calculation are given in the frame. 
}

\figcaption{\label{fig5}
Loci of equi-growth time are traced out in the ($\alpha, \beta$)
plane for the Parker instability under the linear gravity.
}

\figcaption{\label{fig6}
Dispersion relations of the Parker instability under the realistic 
gravity. Each curve is marked by the value of $s$, which is the ratio 
of the velocity dispersion of stars to the effective velocity dispersion 
of clouds. The ordinate represents square of the normalized growth rate, 
and the abscissa the normalized horizontal wavenumber. The system 
parameters are specified within the frame. 
}

\figcaption{\label{fig7}
Dispersion relations of the Parker instability under the realistic 
gravity. Each curve is marked by the vertical wavenumber $\nu_z$. 
The ordinate represents square of the normalized growth rate, and the 
abscissa the normalized horizontal wavenumber. The system  parameters 
are given in the frame.
}

\figcaption{\label{fig8}
Loci of equi-growth time are traced out in the ($\alpha, \beta$) plane 
for the Parker instability under the realistic gravity. 
}

\figcaption{\label{fig9}
Dispersion relations of the uniform, linear, and realistic gravities
are compared with each other. The ordinate represents square of the
normalized growth rate, and the abscissa the normalized horizontal 
wavenumber. The system parameters are specified in the frame.
}

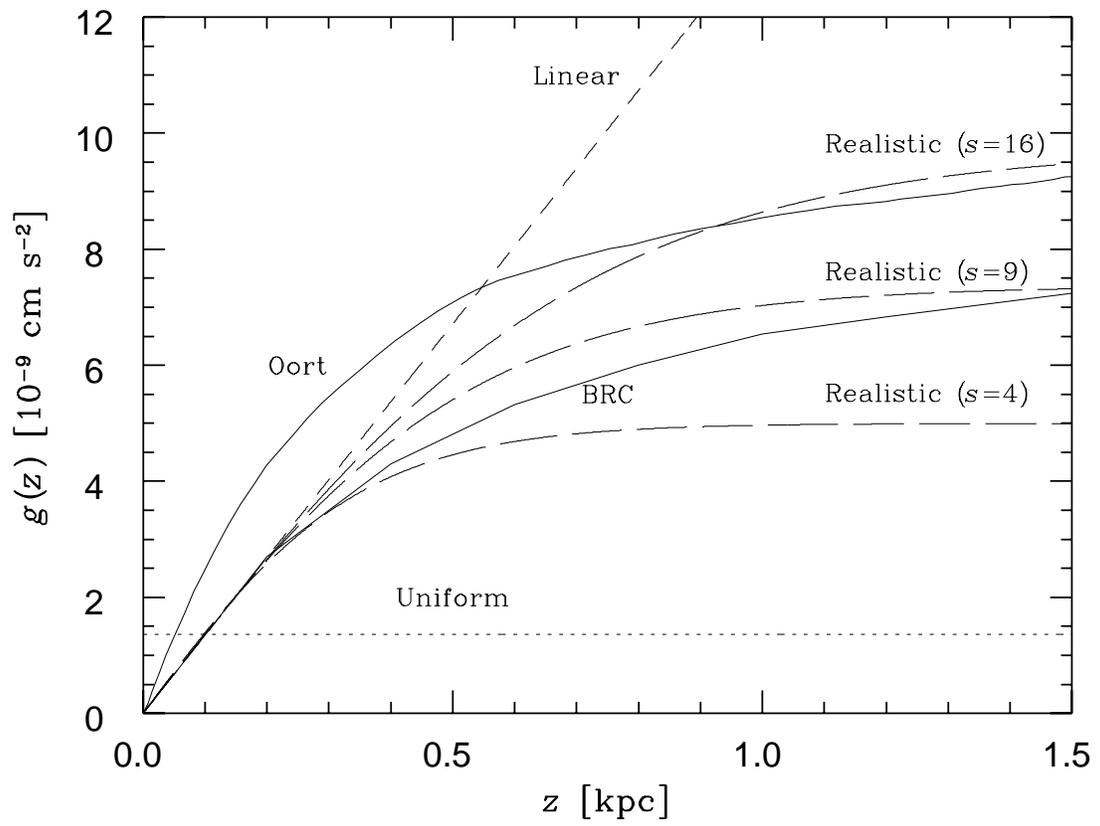
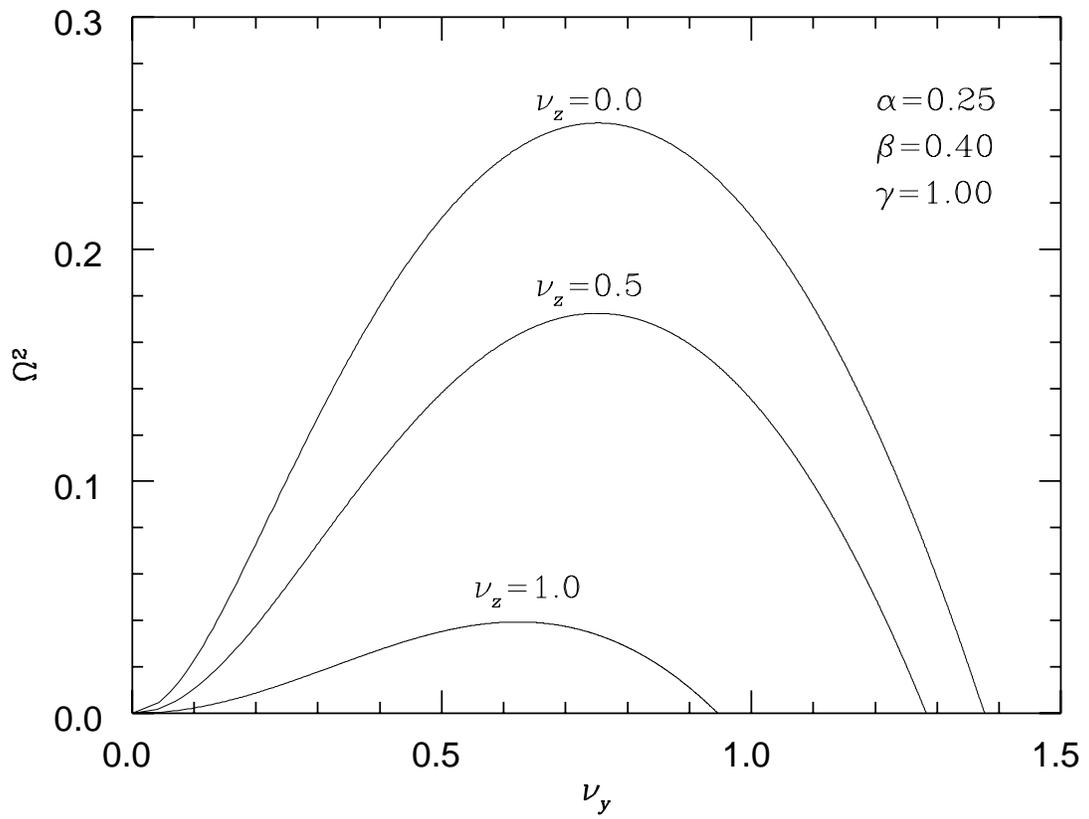
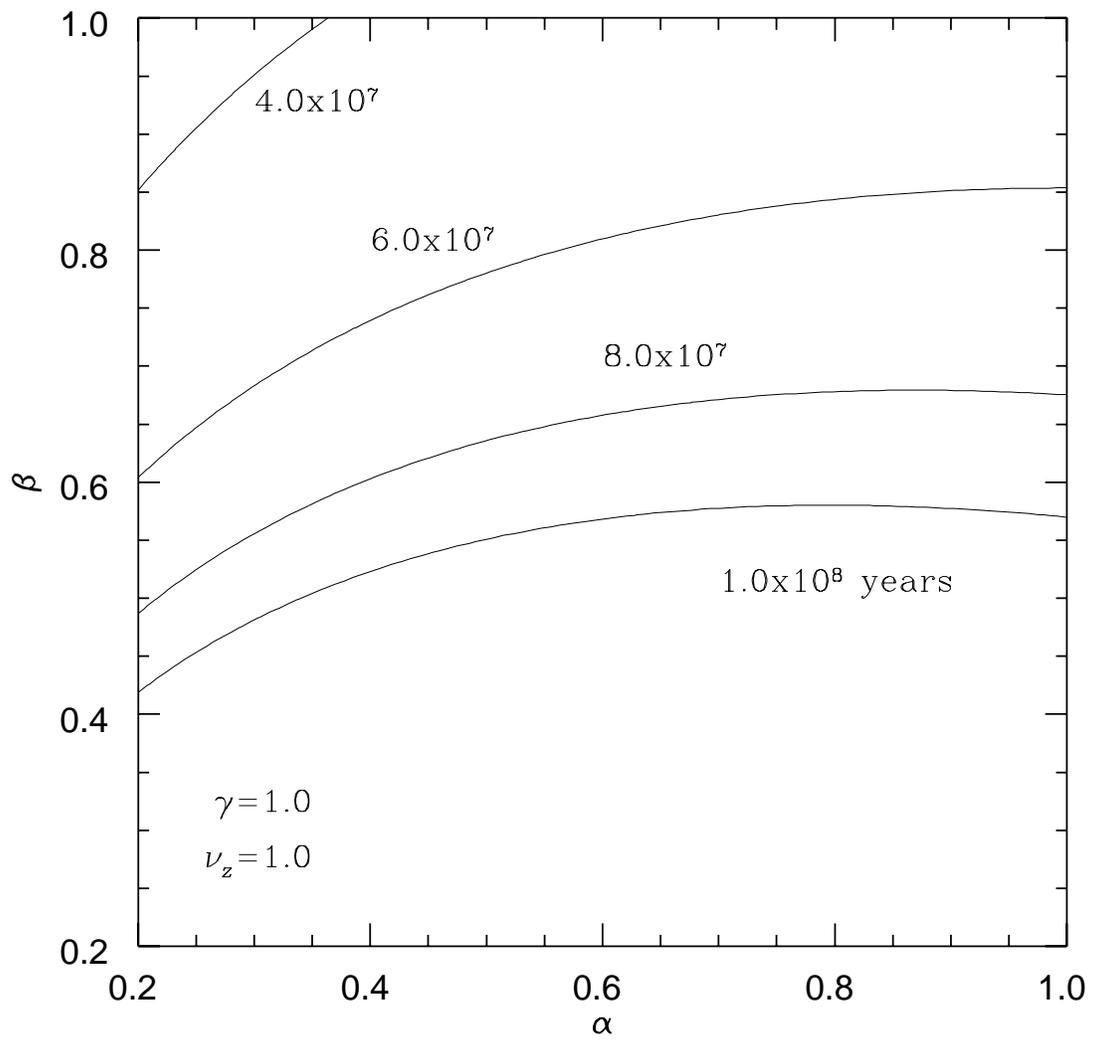
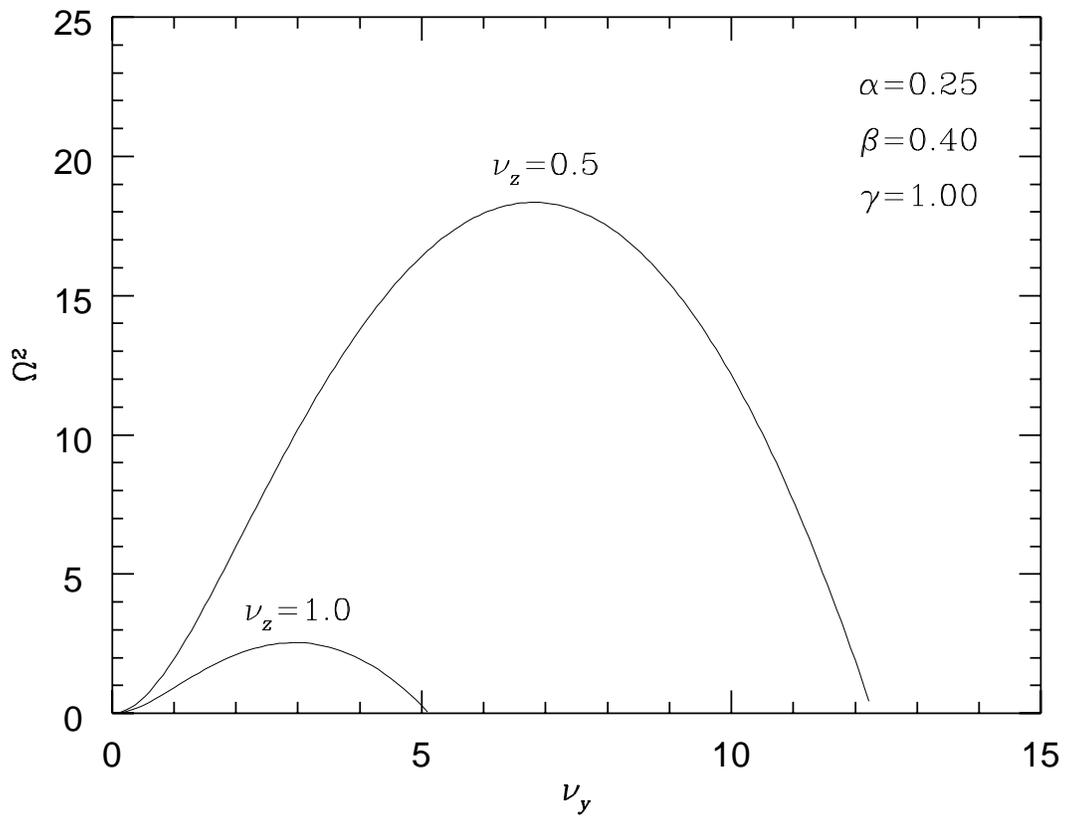
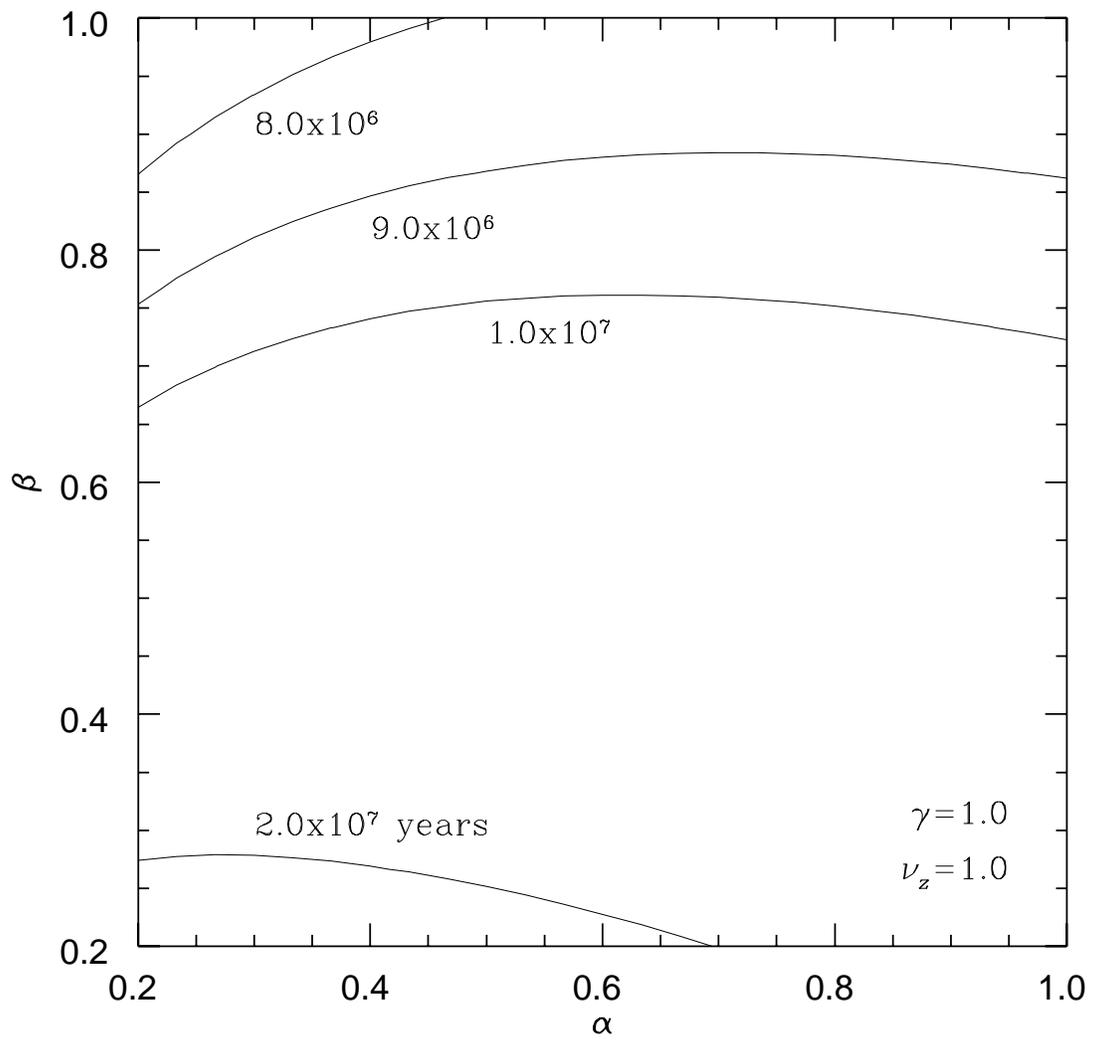
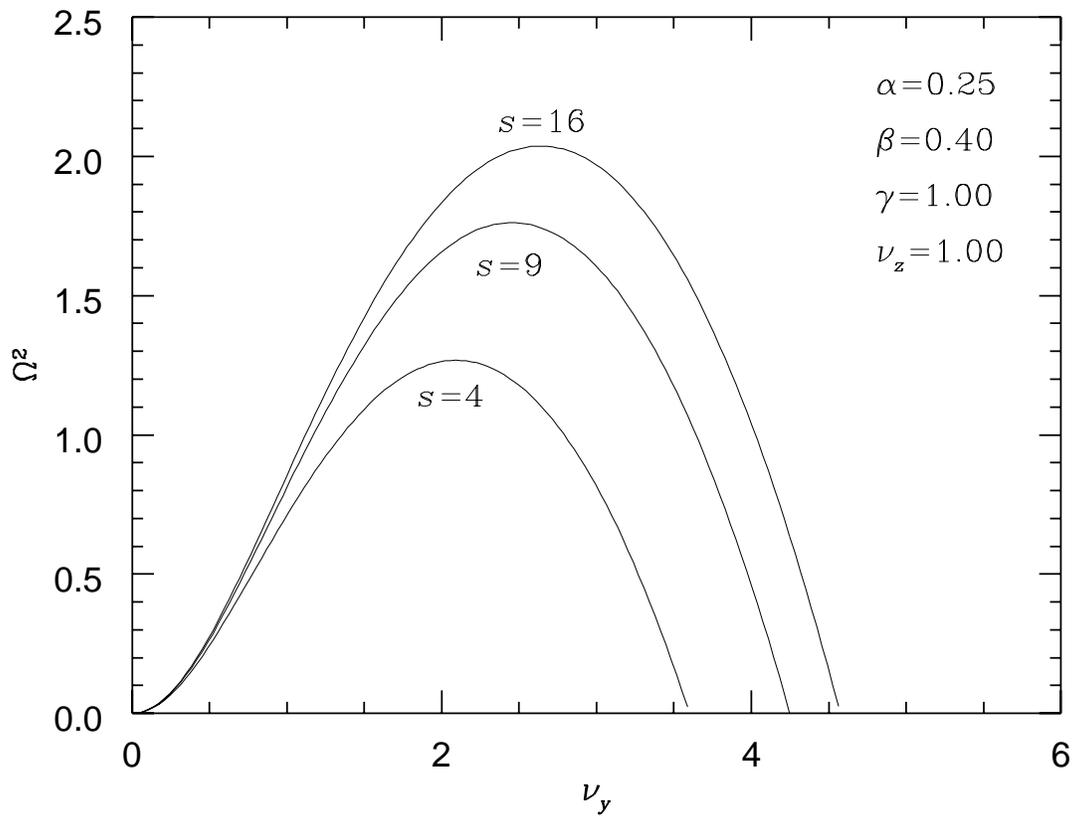
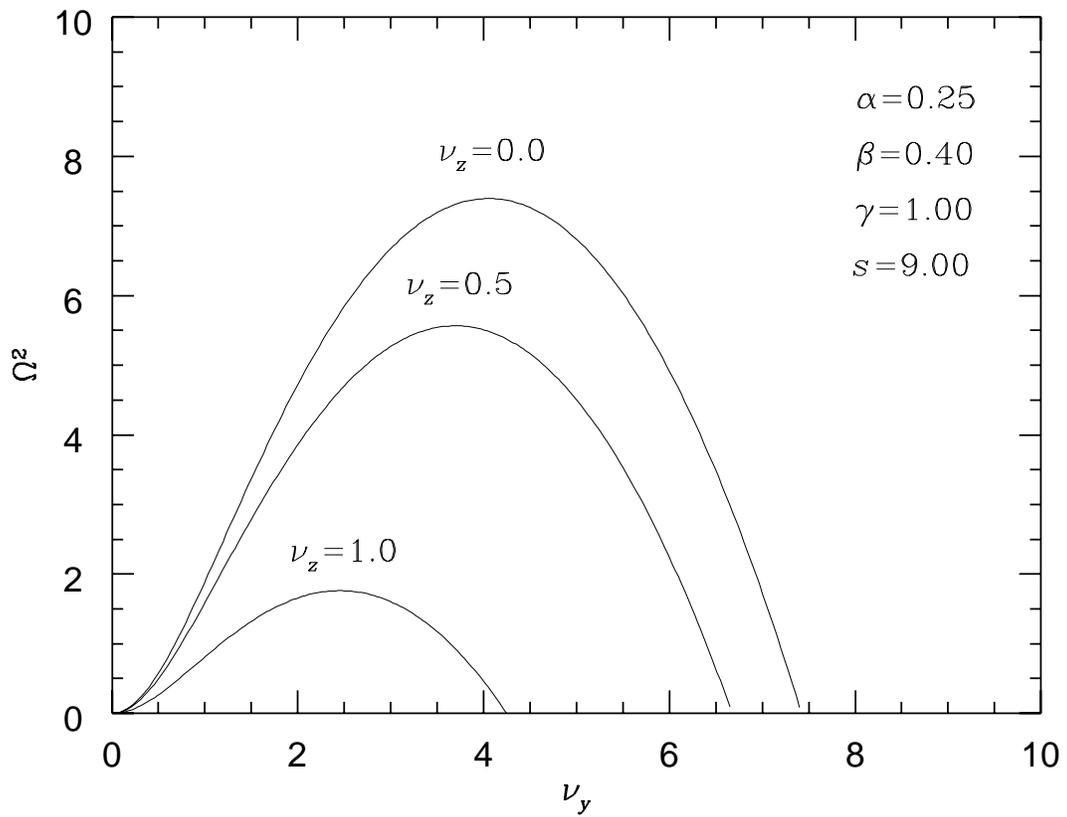
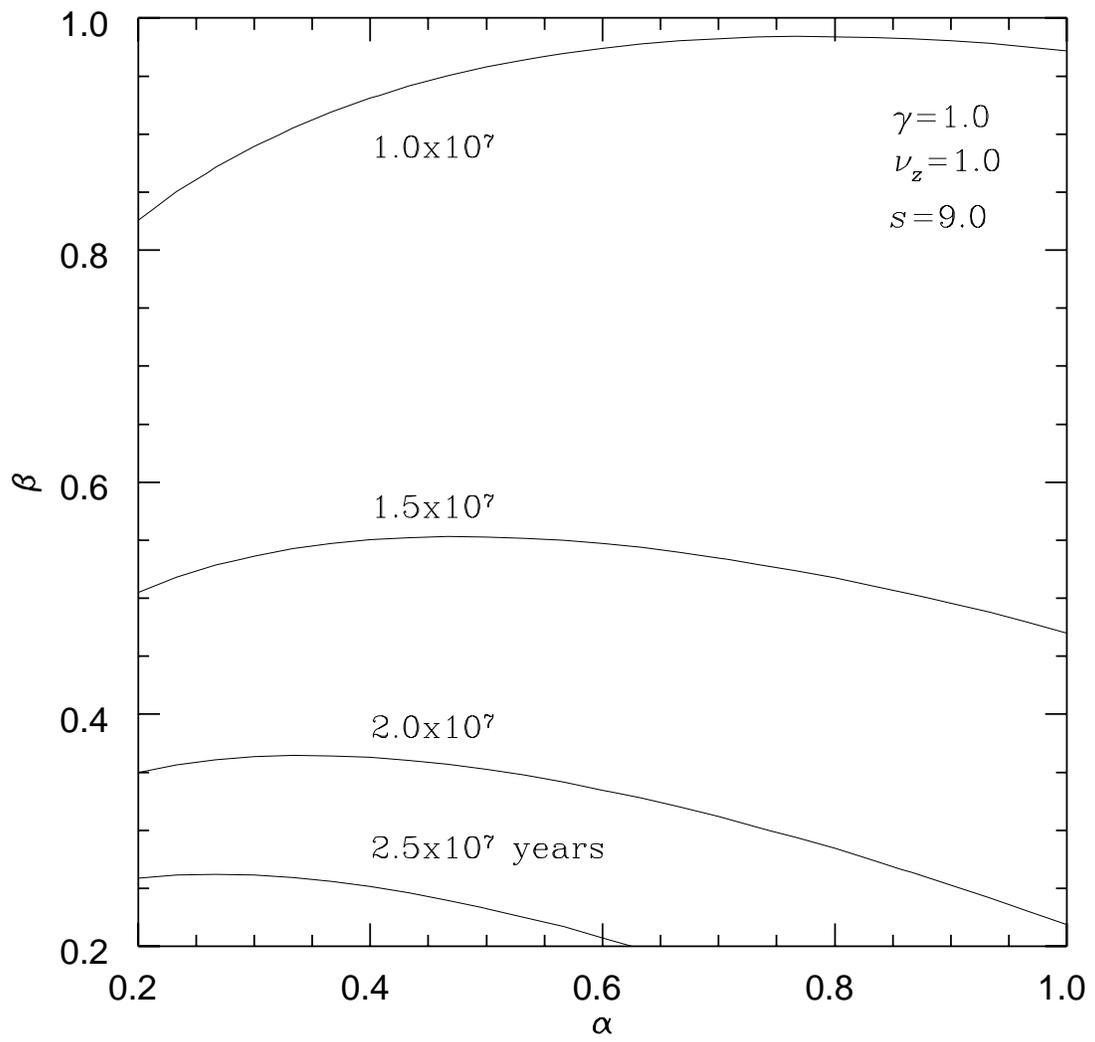
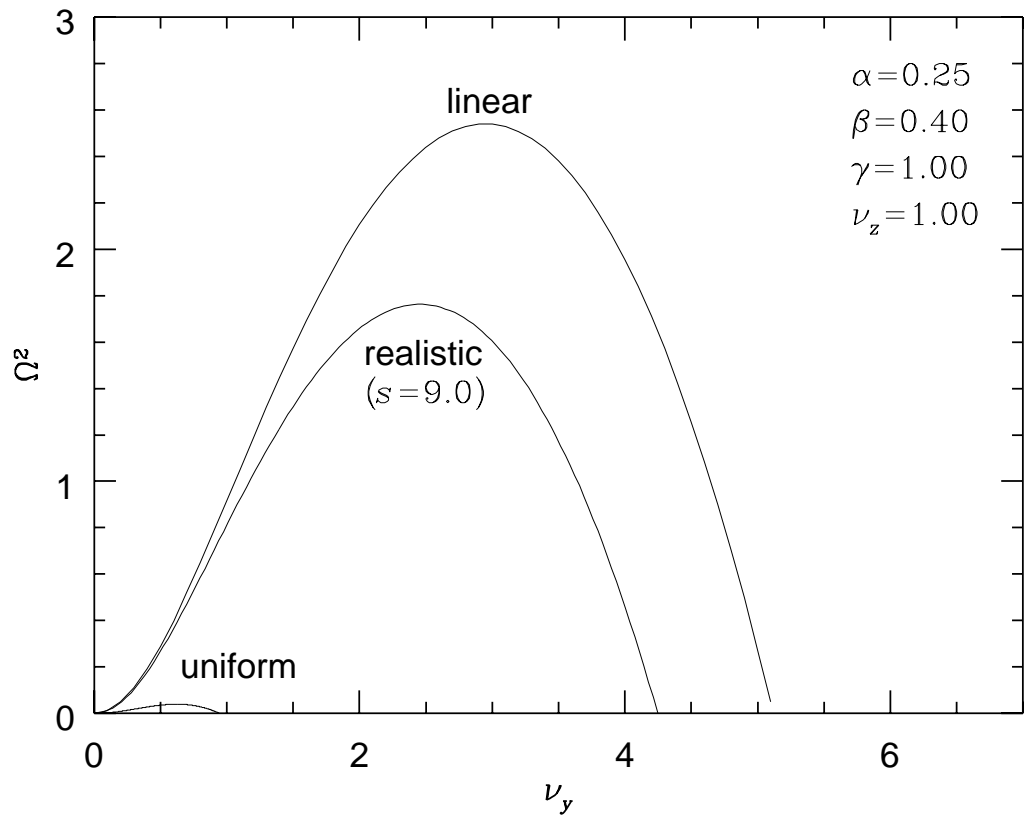
\end{document}